\documentclass[aps,prx,twocolumn,superscriptaddress,floatfix]{revtex4-2}

\usepackage{graphicx}
\usepackage{xcolor}
\usepackage{dcolumn}
\usepackage{bm}
\usepackage{hyperref}
\hypersetup{
    colorlinks=true,
    linkcolor=blue,     
    urlcolor=blue,
    citecolor=blue
}

\newcommand{\Bext}{B_{\mathrm{ext}}}
\newcommand{\Btip}{B_{\mathrm{tip}}}
\newcommand{\Rxx}{R_{\mathrm{xx}}}
\newcommand{\Rxy}{R_{\mathrm{xy}}}
\newcommand{\V}[2]{V_{#1#2}}
\newcommand{\Tc}{T_{\mathrm{c}}}

\begin{document}
\title{Tunable magnetic domains in ferrimagnetic MnSb$_2$Te$_4$}

\author{Tatiana A. Webb}
\affiliation{Department of Physics, Columbia University, New York, NY 10027, USA}
\author{Afrin N. Tamanna}
\affiliation{Department of Physics, The City College of New York, New York, NY 10027, USA}
\author{Xiaxin Ding}
\affiliation{Department of Physics, The City College of New York, New York, NY 10027, USA}
\author{Jikai Xu}
\affiliation{Department of Physics, Columbia University, New York, NY 10027, USA}
\author{Lia Krusin-Elbaum}
\email{krusin@sci.ccny.cuny.edu}
\affiliation{Department of Physics, The City College of New York, New York, NY 10027, USA}
\author{Cory R. Dean}
\email{cd2478@columbia.edu}
\affiliation{Department of Physics, Columbia University, New York, NY 10027, USA}
\author{Dmitri N. Basov}
\email{db3056@columbia.edu}
\affiliation{Department of Physics, Columbia University, New York, NY 10027, USA}
\author{Abhay N. Pasupathy}
\email{apn2108@columbia.edu}
\affiliation{Department of Physics, Columbia University, New York, NY 10027, USA}
\affiliation{Condensed Matter Physics and Materials Science Division, Brookhaven National Laboratory, Upton, New York 11973, USA}


\begin{abstract}
Highly tunable properties make Mn(Bi,Sb)$_2$Te$_4$ a rich playground for exploring the interplay between band topology and magnetism: On one end,  MnBi$_2$Te$_4$ is an antiferromagnetic topological insulator, while the magnetic structure of MnSb$_2$Te$_4$ (MST) can be tuned between antiferromagnetic and ferrimagnetic. Motivated to control electronic properties through real-space magnetic textures, we use magnetic force microscopy (MFM) to image the domains of ferrimagnetic MST. We find that magnetic field tunes between stripe and bubble domain morphologies, raising the possibility of topological spin textures. Moreover, we combine {\textit{in situ}} transport with domain manipulation and imaging to both write MST device properties and directly measure the scaling of the Hall response with domain area. This work demonstrates measurement of the local anomalous Hall response using MFM, and opens the door to reconfigurable domain-based devices in the M(B,S)T family.
\end{abstract}

\maketitle


The recent discovery of MnBi$_2$Te$_4$ (MBT)~\cite{Deng2020, Otrokov2019, Zhang2019, Lei2020, Deng2021} was a breakthrough to realize the quantum anomalous Hall effect in a stoichiometric crystal, avoiding the need for disorder-inducing magnetic dopants~\cite{Lachman2015, Lee2015, Huang2020, Liu2022}. In addition, crystals are exfoliatable down to few layer thicknesses enabling integration into Van der Waals heterostructures with well developed fabrication techniques~\cite{Deng2020, Liu2020, Ovchinnikov2022, Cai2022}. This discovery was rapidly followed by work extending MBT into a family of materials with highly tunable properties via crystallographic and chemical paradigms~\cite{Eremeev2017, Li2019prx, Shi2019, Chen2019, Yan2019, Rienks2019, Murakami2019, Lei2020, Hu2020, Riberolles2021,  Deng2021}. Substituting Sb for Bi changes the doping from n-type to p-type~\cite{Chen2019, Ma2021}. But surprisingly, within MST, the magnetic order can also be tuned (via the concentration of magnetic defects) from A-type antiferromagnetic seen in MBT, where planes of Mn moments are aligned ferromagnetically (antiferromagnetically) within the plane (between planes), to ferrimagnetic with net out of plane magnetization~\cite{Murakami2019, Lai2021, Riberolles2021, Liu2021, Ge2021}. The ability to tune the effective inter-plane coupling from antiferromagnetic to ferromagnetic strongly suggests the presence of magnetic frustration in M(B,S)T, raising the possibility of stabilizing other interesting magnetic orders~\cite{Hayami2016, Li2020}.

The ability to tune magnetic order in the M(B,S)T family opens more conventional applications of magnetic materials, where intense efforts have gone into developing materials structures with interdependent magnetic and electronic properties for control of charge and spin transport (e.g. magnetic data storage and spintronics). Growing evidence suggests that the low energy bands of MST are sensitive to the details of magnetic order~\cite{Murakami2019, Zhang2019, Li2019prb, Wimmer2021, Liu2021, Zhou2020}, but we do not yet have a detailed understanding of the correlation between electronic properties and real-space magnetic textures in MST. So far, the use of magnetism to control electronic properties in magnetic topological materials has been explored primarily in terms of topological phase transitions (e.g.~\cite{Li2019prb, Liu2020, Ovchinnikov2021, Cai2022}) and manipulating chiral edge modes of the quantum anomalous Hall effect~\cite{Yasuda2017, Rosen2017, Ovchinnikov2022}. Edge mode manipulation has been demonstrated via magnetic domains in Cr-doped (Bi,Sb)$_2$Te$_3$~\cite{Yasuda2017, Rosen2017} and via layer-dependent magnetization in antiferromagnetic MBT~\cite{Ovchinnikov2022}, but to the best of our knowledge, the ability to write arbitrary-shaped domains in ferrimagnetic M(B,S)T compounds has yet to be investigated. In this work, we therefore set out to investigate what magnetic textures can be realized in MST, and the prospects for manipulating the local magnetization to create configurable devices. Specifically, we use the domain imaging and writing capabilities of magnetic force microscopy (MFM) combined with {\textit{ in situ}} transport to directly measure the device response to local changes in magnetization.

\begin{figure*}
    \centering
    \includegraphics[width=17.8cm]{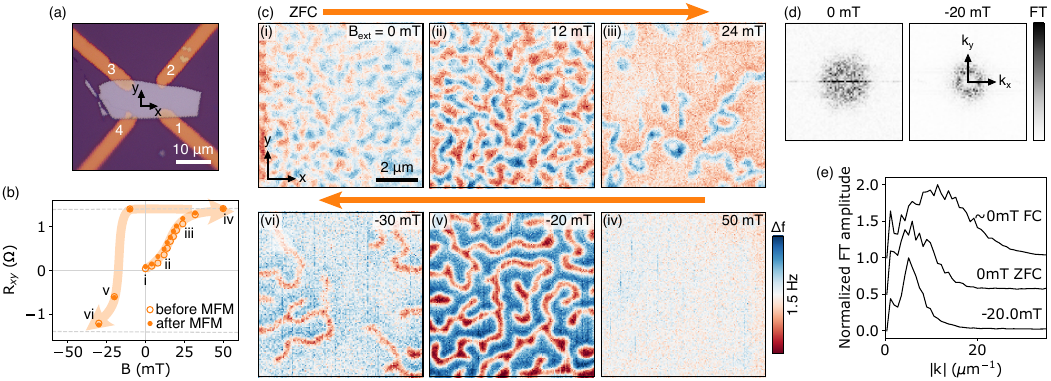}
    \caption{\label{f:stripes} {\bf Evolution of stripe domains under $\Bext$}:
        {\bf (a)} Optical micrograph of the MST device showing the MST flake with 4 contacts for transport measurements in a Van der Pauw geometry. The arrows show the scan axes for the MFM images.
        {\bf (b)} Magnetic field $\Bext$ dependence of the Hall resistance $\Rxy$ measured at 5~K starting from a zero-field cool. Measurements were recorded before (open symbols) and after (filled symbols) MFM imaging. The light orange lines are guides to the eye showing the order of data acquisition.
        {\bf (c)} Constant height MFM measurements of the magnetic domains at the center of the MST device recorded at 5~K after zero-field cooling. Measurements were interspersed with $\Rxy$ and $\Rxx$ measurements shown in (b) and in the Supporting Information SIV. The arrows indicate the order of data acquisition. The color scale on all images is 1.5~Hz, but the zero values have been offset. The tip was lifted 300~nm above the SiO$_2$ surface.
        {\bf (d)} Amplitude of the Fourier transforms of (c-i) and (c-v) after mean value subtraction.
        {\bf (e)} Angular averaged amplitude of the Fourier transformed MFM data.
        }
\end{figure*}

Our interdependent transport and magnetic measurements were performed on an exfoliated flake of ferrimagnetic MST (Figure~\ref{f:stripes}a) with average thickness 84~nm and $\pm10$~nm variations (Supporting Information SII). Four gold contacts were used to measure longitudinal $\Rxx$ and Hall $\Rxy$ resistance.
Immediately after device fabrication, $\Rxx$ showed a peak at 27 K on cooling, consistent with typical Curie temperatures seen in MST. At 2K, the hysteretic loop in $R_{xy}$ and peaks in $R_{xx}$ as a function of magnetic field showed a coercive field near 10~mT (Supporting Information SIV). Refer to Supporting Information SI for additional details of sample fabrication and characterization.


To characterize the magnetic domains in MST, we performed MFM in a cryogenic atomic force microscope (AFM) with variable magnetic field $\Bext$ normal to the sample surface. Figure \ref{f:stripes}c-i shows a MFM image of the zero-field cooled (ZFC) sample at 5~K. Because the coercive field of MST is so low, we quenched the superconducting magnet prior to cooling the sample to ensure a true ZFC with no trapped flux. The MFM image is a measurement of $\Delta f$, the resonance frequency shift of the AFM cantilever due to the interaction of the sample's stray fields with the cantilever's magnetic tip, so we expect images to primarily detect the domain structure of the ferromagnetically aligned components of MST's ferrimagnetic ordering~\cite{Schwarz2008, Hartmann1999}. Correspondingly,  the ZFC image shows disordered maze-like stripe domains (Figure~\ref{f:stripes}) consistent with ferromagnetic ordering in the out of plane direction, similar to domain images from Ge et al~\cite{Ge2021}. Applying magnetic field $B_{ext}$ normal to the sample surface polarizes the sample (c-i to iv), increasing the area of the domains aligned with the field until at 50 mT, only a single domain remains, giving a uniform MFM signal. \textit{In situ} transport measurements show an associated increase in $\Rxy$ from 0.04 to 1.42 $\Omega$. 
Over this range of $\Bext$, the contribution to $\Rxy$ from the linear Hall effect is negligible, so the change in $\Rxy$ is primarily due to the anomalous Hall Effect (AHE)~\cite{Nagaosa2010} (Supporting Information SVII).

Reversing the magnetic field (Figure~\ref{f:stripes}c-iv to vi), we observe the reformation of stripe domains at -20 mT as $\Rxy$ drops and changes sign, indicating the reversal of the magnetization. These field-reversed domains are significantly less disordered than the ZFC domains. To quantify the difference, we examine the Fourier Transforms (FT) of the ZFC and -20~mT images, shown in Figure~\ref{f:stripes}d. Both exhibit a ring shape, or a peak in the angular-averaged FT (Figure~\ref{f:stripes}e), indicating the domains have a characteristic length scale, as expected from the energetics of domain formation~\cite{Kittel1949, Giess1980, Seul1995}. The peak occurs at wavevector $|k|$ 6.3~$\mu m^{-1}$ with standard deviation $\sigma$ 4.8~$\mu m^{-1}$ for the ZFC domains and $|k|=$ 5.7~$\mu m^{-1}$  with $\sigma=$ 2.1~$\mu m^{-1}$ for the -20 mT domains. The broader peak associated with the ZFC domains indicates that during cooling the domains form features with a wider range of length scales compared to during magnetization reversal at low temperature.


\begin{figure*}
    \centering
    \includegraphics[width=17.8cm]{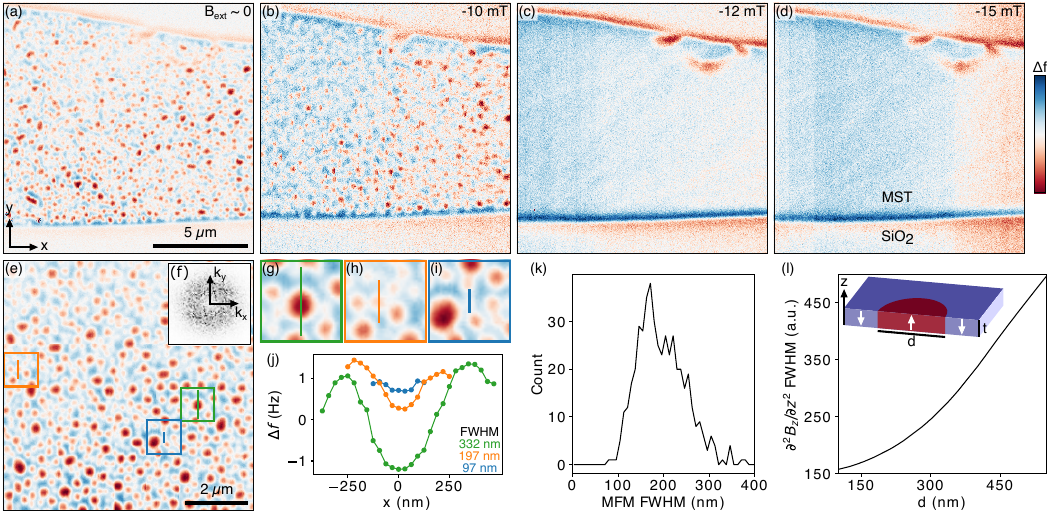}
    \caption{\label{f:bubbles} {\bf Field cooled domain structures.}
    {\bf (a-d)} Constant height MFM images of the magnetic domains in the MST device under field cooling with the indicated $\Bext$. The tip was lifted 300~nm above the SiO$_2$ surface. The range and offset of the color scale has been chosen for each image independently. Color scale range: 3.7~Hz (a), 1.5~Hz (b), 1.4~Hz (c), 1.5~Hz (d). Temperature: 5~K (a), 10~K (b-d).
    {\bf (a)} $\Bext \sim 0$ indicates that a small unknown residual flux from the superconducting magnet was present.
    {\bf (e)} Smaller scale constant height MFM image after cooling under $\Bext \sim 0$ showing clear bubble shaped domains. Color scale range: 4.4~Hz. Temperature: 5~K. The images in (a) and (e) are from separate cooling cycles with the same $\Bext$.
    {\bf (f)} Amplitude of the Fourier transform of (e) after mean value subtraction.
    {\bf (g-i)} Zooms of 3 regions in (e) showing a large, medium, and small size bubble.
    {\bf (j)} Profiles through the large, medium, and small size bubbles from (g-i).
    {\bf (k)} Histogram of the full-width-at-half-max (FWHM) of the bubbles imaged via MFM, determined from horizontal and vertical profiles through all resolvable bubbles in (e). The MFM FWHM is not directly interpretable as the bubble domain size.
    {\bf (l)} FWHM of $\partial^2 B_z / \partial z^2$ for the stray magnetic field generated by a bubble domain with diameter d, using sample thickness $t=$~81.2~nm and $z=$~300~nm measured from the bottom of the material.
    }
\end{figure*}

To further explore how an external field can tune the domain morphology, we cooled the MST device from 35~K below $\Tc$ under $|\Bext|$ up to 15~mT (Figure~\ref{f:bubbles}).
With $|\Bext|$ larger than 10 mT, a single domain forms across the entire MST flake.
However, when we nominally zero the magnet's current such that a small $\Bext \approx 0$ exists only from trapped flux, we see circular features in the MFM, indicating the formation of bubble rather than stripe domains. MST is thus remarkably sensitive to small magnetic fields. At intermediate $|\Bext|$ (10 mT), the domains formed are not uniform in size and shape, and it is not clear if they are intrinsically bubbles or stripes. We now focus on $\Bext \approx 0$, where bubbles are clearly observed.

The bubble domains are highly disordered (Figure~\ref{f:bubbles}e). The nearly isotropic Fourier transform (f) shows no evidence for lattice organization, and the distribution of wavevectors centered at $|k| =$~11.3~$\mu m^{-1}$ with $\sigma=$~7~$\mu m^{-1}$ is extremely broad. Correspondingly, the circular MFM features range in size from below 100~nm to above 300~nm, as shown by the distribution of full width at half maxima (FWHM, Figure~\ref{f:bubbles}k). 
The size of the features seen in MFM cannot directly be interpreted as the size of the domains in the MST. Approximating the tip as a point dipole with small oscillation amplitude, the MFM image $\Delta f$ is proportional to $\partial^2 B_z / \partial z^2$ arising from the sample's stray field~\cite{Schwarz2008, Hartmann1999}. To help interpret the MFM features, we model the stray field for a single cylindrical bubble domain at a representative height $z$. 
As the domain diameter decreases below $z$, the spatial peak in $\partial^2 B_z / \partial z^2$ decreases in intensity ( Supporting Information SV) and the FWHM saturates at a lower limit near 150~nm (Figure~\ref{f:bubbles}l). The FWHM does not decrease linearly in the domain diameter for small bubbles. Returning to the MFM data, we therefore expect small bubbles may not be detectable due to weak intensity, and for slightly larger bubbles, the apparent size in MFM may saturate at a lower limit larger than the domain diameter. The MFM data, however shows bubbles with FWHM below the expected 150~nm cuttoff, likely because the width of the low intensity bubbles can be dominated by the positions of the neighboring bubbles. Under repeat cooling, we find that some but not all bubbles form in the same location (Supporting Information SIX), which along with their disordered organization could indicate significant pinning, either due to crystal inhomogeneity, or extrinsic factors such as strain. The observation of bubble domains under field cooling, but not when sweeping $\Bext$ at low temperature suggests that the bubble and stripe morphologies are separated by a significant energy barrier, likely associated with nucleating a domain wall.

The domain wall structure (i.e. Bloch or N\'eel) can produce topologically non-trivial chiral spin textures on bubble domains~\cite{Malozemoff1979}, and topological bubble and skyrmion phases have been reported in other (Bi,Sb)$_2$Te$_3$-based materials~\cite{Jiang2020, Wang2021}. While our detection of bubble domains opens the possibility of stabilizing topological spin textures in M(B,S)T, our MFM measurements do not allow us to draw a conclusion about the topology of the bubbles. We observed no evidence of a topological Hall effect (THE) -- a deflection of carriers due to the real space Berry curvature of topological spin textures -- in the $\Rxy$ hysteresis loops when sweeping $\Bext$ to flip the sample magnetization at low temperature. However, unlike many skyrmion materials that display a THE as the skyrmion phase is formed over a finite range of $B$ at constant temperature, the MFM data in Figure~\ref{f:stripes} does not show the bubble morphology when sweeping $\Bext$ at low temperature. Further work is therefore required to determine if the bubble domains formed on field cooling are topological or trivial.


\begin{figure}
\centering
    \includegraphics[width=8.46cm]{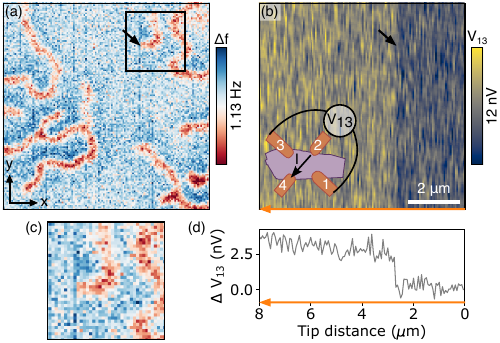}
    \caption{\label{f:domain-flip} {\bf Impact of a single domain on transport}
    {\bf (a)} MFM image showing a domain flipping during the scan, indicated by the black arrow. Same as Figure~\ref{f:stripes}(c-vi).
    {\bf (b)} $\V13$ measured simultaneously with (a), using a 500~nA amplitude AC source current. The black arrow indicates the location of the domain flip, identical to the arrow in (a). Inset: Schematic of the $\V13$ measurement.
    {\bf (c)} Zoom of (a) on the area showing the domain flip. The fast scan direction is vertical. The domain abruptly disappears from one vertical scan line to the next.
    {\bf (d)} $\V13$ averaged vertically along the fast scan direction to show the jump in value that occurred as the domain flipped. The value from the first scan line has been subtracted to show the change $\Delta \V13$.
    }
\end{figure}

We have seen how the domain morphology can be controlled with $\Bext$; now we investigate the possibility of using the stray field from the magnetic MFM tip, $\Btip$, to locally manipulate the domains in MST. To reduce the influence of $\Btip$ on the sample, the domain imaging discussed so far was done with the tip lifted high (roughly 200-230~nm) above the MST surface.
However, when $\Bext$ is near the coercive field, small changes in the magnetic field can have a large influence on the sample magnetization, and even 200 nm from the tip, $\Btip$ could be on the order of 10~mT~\cite{Rizzo2022}, comparable to the coercive field.
Correspondingly, small changes in $R_{xy}$ and $R_{xx}$ during MFM imaging (Figure~\ref{f:stripes}b, Supporting Information SIV) demonstrate that the tip mildly influenced the sample magnetization. Moreover, tip-induced domain flips are seen in some images as a domain that abruptly disappears partway through imaging.
To quantify the tip's influence, we applied an AC current between contacts 2 and 4, and measured the induced transverse voltage $\V13$ across contacts 1 and 3 during MFM imaging.
The MFM image in Figure~\ref{f:domain-flip}a shows a domain flip, and the simultaneously acquired $\V13$ image (b,d) shows an abrupt change by more than 2 nV at the same location, demonstrating a measurable impact of the domain flip on the device transport.
Interestingly the tip-induced domain flips are not always in the sense of aligning the domain with the tip, suggesting that the spatial gradient or time-dependence of $\Btip$ may be equally important or more important for overcoming energy barriers compared to the local Zeeman energy term.

\begin{figure*}
    \centering
    \includegraphics[width=17.8cm]{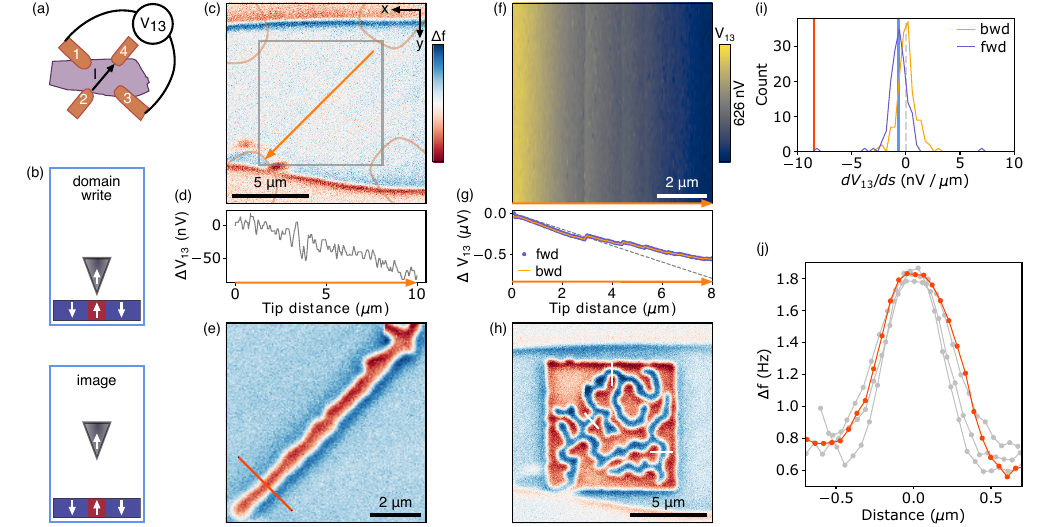}
    \caption{\label{f:domain-write} {\bf Writing magnetic domains.}
    {\bf (a)} Schematic of the device and the $\V13$ measurement used as an approximation of the Hall response during domain writing.
    {\bf (b)} Schematic of the magnetic tip over the MST sample during domain writing, with the tip close to the surface, and during domain imaging, with the tip lifted high above the surface.
    {\bf (c)} MFM showing the MST device has no domains, and has magnetization anti-aligned with the tip after ramping $\Bext$ to -50~mT and then 0~mT. Constant height imaging was done with the tip lifted 300~nm above the SiO$_2$ surface.
    The overlays show the approximate locations of: the electrical contacts (orange lines), the tip trajectory for writing the line domain (orange arrow), and the MFM image of the line domain (gray box).
    Color scale range: 1.1~Hz.
    {\bf (d)} $\V13$ measured while writing the line domain. 500~nA amplitude AC source current.
    {\bf (e)} Constant height MFM image after writing the line domain, with the tip lifted approximately 200~nm above the MST surface. The red line indicates the location of the cut shown in (j).
    Color scale range: 1.9~Hz.
    {\bf (f)} $\V13$ recorded as a function of the tip position while attempting to write a square area. See main text. 500~nA amplitude AC source current.
    {\bf (g)} $\V13$ from (f) averaged vertically along the fast scan direction. The value of the first point was subtracted to show the change $\Delta\V13$. The gray dashed line is the expected linear trend if the tip were fully polarizing the sample.
    {\bf (h)} Constant height MFM image after writing the square area. Tip lifted 300~nm above the SiO$_2$ surface. The white lines indicate the locations of the cuts shown in (j).
    Color scale range: 1.6~Hz.
    {\bf (i)} Histograms of the slopes of $\V13$ during each forward and backward vertical scan line while writing the square area (f). The vertical lines mark the slope while writing the line domain (red), and the expected per-pixel slope (blue) calculated from the overall change in $\V13$ and the domain area in (h).
    {\bf (j)} Line cuts through the MFM images of the line domain (red) and of the square area (gray). All cuts have been offset for comparison, and the cuts from the square domain have been inverted.
    Temperature: 10~K.
    }
\end{figure*}

We can harness the tip's influence to controllably write domains by bringing the tip close to the MST surface, increasing $\Btip$. For this purpose, we first used $\Bext$ to prepare the sample with magnetization anti-aligned to the tip (Figure~\ref{f:domain-write}c). After zeroing $\Bext$, we then brought the tip into amplitude-controlled feedback on the MST surface ($\Btip$ on the order of 50 mT~\cite{Rizzo2022}), and moved the tip across the surface to write a domain aligned with the tip. In Figure~\ref{f:domain-write}, we show both linear (e) and square (h) areas written with the MFM tip, demonstrating that both narrow 1D-like and 2D domains can be written. During the write process, the square area formed a mixed domain state, suggesting that because the mixed domain state is energetically favored at $\Bext=0$, there is a maximum single-domain area of roughly several $\mu$m$^2$ (Supporting Information SXI) that can be written. Decreasing temperature to increase the importance of the domain wall nucleation energy may increase that area.

By inverting the magnetization of a small area locally with our MFM tip, we can directly probe that area's impact on the AHE. During domain writing we therefore recorded $\V13$ as a proxy for $\Rxy$ (Supporting Information SVIII). While writing the line domain, $\V13$ decreased linearly (Figure~\ref{f:domain-write}d), matching the area-scaling that one would expect for AHE contributions~\cite{Nagaosa2010} that scale linearly with the sample average magnetization. $\V13$ recorded while writing the square domain is also consistent with area scaling. Here, $\V13$ forms two 2D images for forward (Figure~\ref{f:domain-write}f) and backward scans--to write the domain, the tip moved up and down along each scan line before advancing one pixel at a time left to right. Typically, $\V13$ has a finite slope on the forward pass (the blue histogram in i is peaked at 0.6~nV$/\mu$m), confirming that the tip is writing a magnetization, but not on the backward pass (orange histogram, peaked at zero). Considering that typically each scan line advances the domain wall by one pixel width (53~nm), we can quantify the local AHE: 11 nV/$(\mu\mathrm{m})^2$. This value is quantitatively consistent with both: (1) the linear $\V13$ seen when writing the line domain, and (2) the ratio of the change in $\V13$ from before to after the write step to the domain area imaged via MFM (Supporting Information SXI). Moreover, the entire evolution of $\V13$ during the square write can be understood in detail as a linear decrease (gray dashed line in g) from writing the red domain plus deviations from forming the inner blue domain, in abrupt steps initially but then more smoothly near the end of the write. 

We have therefore demonstrated a direct measurement of the scaling of the anomalous Hall effect with domain area. The technique can also in principle measure deviations from this area scaling to probe local properties of inhomogenous devices (spatially varying magnetization or Berry curvature). Within homogenous materials, the area-scaling contributions and deviations represent bulk and boundary contributions, meaning that this technique can be used to probe topological effects such as dissipationless chiral edge conduction in a Chern insulator or the topological Hall effect from chiral spin textures at domain walls.

This work opens the door to making programmable magnetic devices within  ferrimagnetic compounds in the M(B,S)T family. M(B,S)T could be a platform for writable chiral currents (e.g.~\cite{Yasuda2017, Rosen2017, Ovchinnikov2022}) either in a magnetic Weyl semimetal or Chern insulating state (multiple potential band topologies have been predicted in MST~\cite{Chen2019, Zhang2019, Murakami2019, Zhou2020, Wimmer2021, Ma2021}). The ability to tune magnetic domains in a compound that retains magnetic order when exfoliated to the few layer limit~\cite{Chen2019, Deng2020, Ovchinnikov2021, Zang2022} raises the possibility of using M(B,S)T to introduce programmable magnetic landscapes (e.g. supperlattices or boundaries made of magnetic gradients) on length scales of 100s of nanometers to micrometers into van der Waals heterostructures. Generically, the tip writing process allows us to locally move between different metastable magnetic configurations that are separated by energetic barriers. So beyond writing individual domains of uniform magnetization explicitly, the tip influence could be combined with external fields and temperature to stabilize and write areas of non-uniform spin textures (just as the mixed domain state formed in our square area was not uniform) in order to create functional devices based on boundaries between magnetic phases.

\section*{Supporting Information}
Additional experimental details, characterization of the sample topography, images of stripe domains, electrical transport measurements, analysis and modelling of domain length scales, analysis of the repeatability of bubble domain locations, and analysis of the scaling of the AHE with domain area (PDF)

\begin{acknowledgments}
We thank Zachariah Addison and Nishchhal Verma for helpful discussions. This work was supported by the Air Force Office of Scientific Research via grant FA9550-21-1-0378 (T.A.W., A.N.P.) and by NSF grants DMR-2210186 (D.N.B) and HRD-2112550 (L.K.-E.). Research on topological properties of moiré superlattices is supported as part of Programmable Quantum Materials, an Energy Frontier Research Center funded by the U.S. Department of Energy (DOE), Office of Science, Basic Energy Sciences (BES), under award DE-SC0019443. Sample synthesis is supported by the the NSF MRSEC program through Columbia University in the Center for Precision-Assembled Quantum Materials under award number DMR-2011738.
\end{acknowledgments}

\bibliography{mst-domains}

\end{document}


\title{Supporting information for\\ Tunable magnetic domains in ferrimagnetic MnSb$_2$Te$_4$}


\author{Tatiana A. Webb}
\affiliation{Department of Physics, Columbia University, New York, NY 10027, USA}
\author{Afrin N. Tamanna}
\affiliation{Department of Physics, The City College of New York, New York, NY 10027, USA}
\author{Xiaxin Ding}
\affiliation{Department of Physics, The City College of New York, New York, NY 10027, USA}
\author{Jikai Xu}
\affiliation{Department of Physics, Columbia University, New York, NY 10027, USA}
\author{Lia Krusin-Elbaum}
\affiliation{Department of Physics, The City College of New York, New York, NY 10027, USA}
\author{Cory R. Dean}
\affiliation{Department of Physics, Columbia University, New York, NY 10027, USA}
\author{Dmitri N. Basov}
\affiliation{Department of Physics, Columbia University, New York, NY 10027, USA}
\author{Abhay N. Pasupathy}
\affiliation{Department of Physics, Columbia University, New York, NY 10027, USA}
\affiliation{Condensed Matter Physics and Materials Science Division, Brookhaven National Laboratory, Upton, New York 11973, USA}

\maketitle

\tableofcontents
\clearpage

\section{Methods}

\subsection{Crystal growth and structural characterization. }

Crystals of nominally MnSb$_2$Te$_4$ were grown out of a Sb–Te flux~\cite{Deng2022, Yan2019}. Mixtures of Mn (Alfa Aesar, 99.99\%) and Sb pieces (Alfa Aesar, 99.999\%), and Te shot (Alfa Aesar, 99.9999\%) in the molar ratio of 1:10:16 (MnTe:Sb2Te3 = 1:5) were placed in a 2~ml alumina growth crucible and heated to 900~C and held for 12~h. After slowly cooling across a $\sim$10 degree window below 600~C in two weeks, the excess flux was removed by centrifugation above the melting temperature of Sb$_2$Te$_3$ ($\geq$~620~C). Crystals produced by this flux method were typically a few mm on a side and often grew in thick, block-like forms with thicknesses up to 2~mm but were easily delaminated.

{\bf EDX:} Energy Dispersive X-ray (EDX) microanalysis was performed in the Zeiss Supra 55, a field emission SEM with a maximum resolution of 1~nm. The ferrimagnetic (FM) MST stoichiometry was determined as Mn:Sb:Te$=$1.3 : 2.9 : 5.8. {\bf XRD:} X-ray diffraction of crystals was performed in a Panalytical diffractometer using Cu Ka ($\lambda=1.5405$~\AA) line from Philips high intensity ceramic sealed tube (3~kW) X-ray source with a Soller slit (0.04~rad) incident and diffracted beam optics. The determined c-axis parameter 40.898~\AA was consistent with space group, R-3m as reported in literature. {\bf Magnetization:} D.~c. magnetization measurements were performed using Quantum Design SQUID Magnetometer in up to 5.5~T fields. In all magnetic measurements the samples were supported in gelcaps without any substrates. From the fits to Curie-Weiss law, the magnetic moment in as-grown crystals was determined as $\mu_{\mathrm{eff}} = 5.36 \mu_{\mathrm{B}}/\mathrm{mole}$.

\subsection{Magnetic force microscopy}

Magnetic force microscopy (MFM) measurements were performed in an Attocube cantilever-based cryogenic atomic force microscope (attoAFM I with attoLIQUID 2000 cryostat), where the microscope sits in a helium exchange gas at low temperature. Nanosensors PPP-MFMR probes with hard magnetic coating, resonant frequencies near 75~kHz, and force constants near 2.8~N$/$m were used for all measurements. All low temperature measurements on the MST device were were performed using a single AFM probe with cantilever resonance at 78~kHz. 
MFM images record the resonant frequency shift of the probe cantilever ($\Delta f$).
Constant height MFM images were taken with the tip at a fixed height above the plane of the SiO$_2$ surface. 
Constant lift MFM images were taken by passing twice over each line. The first pass, in amplitude-controlled feedback, recorded the surface topography. On the second pass, the tip was lifted by a constant offset from the topographic pass to record $\Delta f$. In this way, topographic and MFM images were recorded over the same field of view in a line-by-line interleaved fashion. During MFM imaging, the tip oscillation amplitude was typically 35~nm to 55~nm. MFM data shown is raw data unless otherwise noted. 

\subsection{Electrical transport measurements}

\textit{Ex situ} transport measurements were performed in a 14 Tesla Quantum Design Physical property measurement system (PPMS) in 1~Torr (at low temperature) of He gas. Crystals were mechanically exfoliated onto 285~nm SiO$_2$/Si wafers. Electrical contacts in the van der Pauw (vdP) configuration were photo-lithographically patterned and a sputtered Au metallurgy was used. Conformal Au coating amply covered side surfaces in order to make good contacts to top and bottom surfaces. The vdP DC measurements were carried out using a custom-configured electronic system in which four measurement configurations are switched by a Keithley scanner, with the current direction reversal employed for each measurement to minimize thermal emf.

\textit{In situ} electrical tranport measurements were done using a Signal Recovery model 7265 lock-in amplifier to source a 17.777 Hz AC voltage, and measure the response voltage at the same frequency. A 10~M$\Omega$ resistor was used to convert the voltage source into a current source. According to the vdP technique~\cite{Vanderpauw1958}, $\Rxx$ was calculated from four measurements, 
$V_{23, 14}$, $V_{12, 43}$, $V_{41, 32}$, and $V_{34, 21}$, and $\Rxy$ from two measurements, $V_{31, 24}$ and $V_{24, 13}$, 
where $V_{ab, cd}$ is the voltage measured from c to d while applying the source current from a to b, and the contacts are numbered in Figure~1a of the main text. For brevity, we refer to $V_{31, 24}$ and $V_{24, 13}$ as $V_{24}$ and $V_{13}$, respectively.

\subsection{Modeling bubble domains}
Calculations of the stray magnetic field arising from a bubble domain were done using Magpylib~\cite{magpylib2020}.
The sample model consisted of a rectangular slab of uniform magnetization with a cylinder of the same thickness located at the center. The cylinder had twice the magnetization with the opposite sign, such that the net magnetization inside the cylinder was equal to that of the rectangular slab, but opposite in direction. The lateral size of the rectangular slab was taken to be 100~mm in both directions, large enough to avoid edge effects in the vicinity of the bubble domain. After calculating the stray magnetic field above the sample surface, numerical differentiation was used to calculate the stray field gradients.
    
\clearpage
\section{Sample topography}

\begin{figure}[h]
    \centering
    \includegraphics[width=17.78cm]{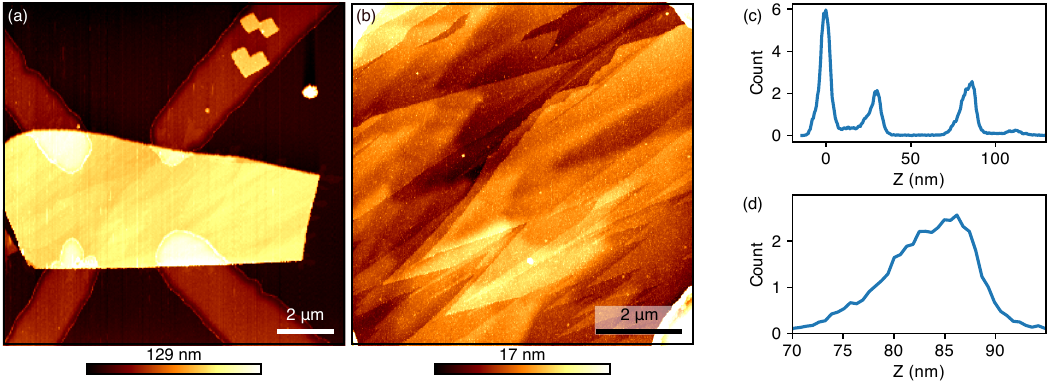}
    \caption{\label{f:topography}
        {\bf (a, b)} AFM topographies (tapping mode) of the full MST device at ambient (a) conditions, and of the center of the device, with the 4 contacts at the corners of the image, at 35~K (b). Background subtraction: Line-by-line constant subtraction based on the median of the difference between consecutive lines, followed by plane subtraction to level the image.
        {\bf (c)} Histogram of (a), showing 4 peaks associated with the SiO$_2$ surface, the gold contacts on SiO$_2$, the MST surface, and the gold contacts on MST.
        {\bf (d)} Zoom of (c) showing just the peak from the MST surface. For the MST surface only, the mean height is 83~nm with standard deviation 4 nm. Considering only the center of the MST device, where most MFM imaging was performed, the mean height is 81~nm with 4~nm standard deviation.
        }
\end{figure}
\clearpage

\section{Evolution of stripe domains under $\Bext$}

\begin{figure}[h]
    \centering
    \includegraphics[width=17.78cm]{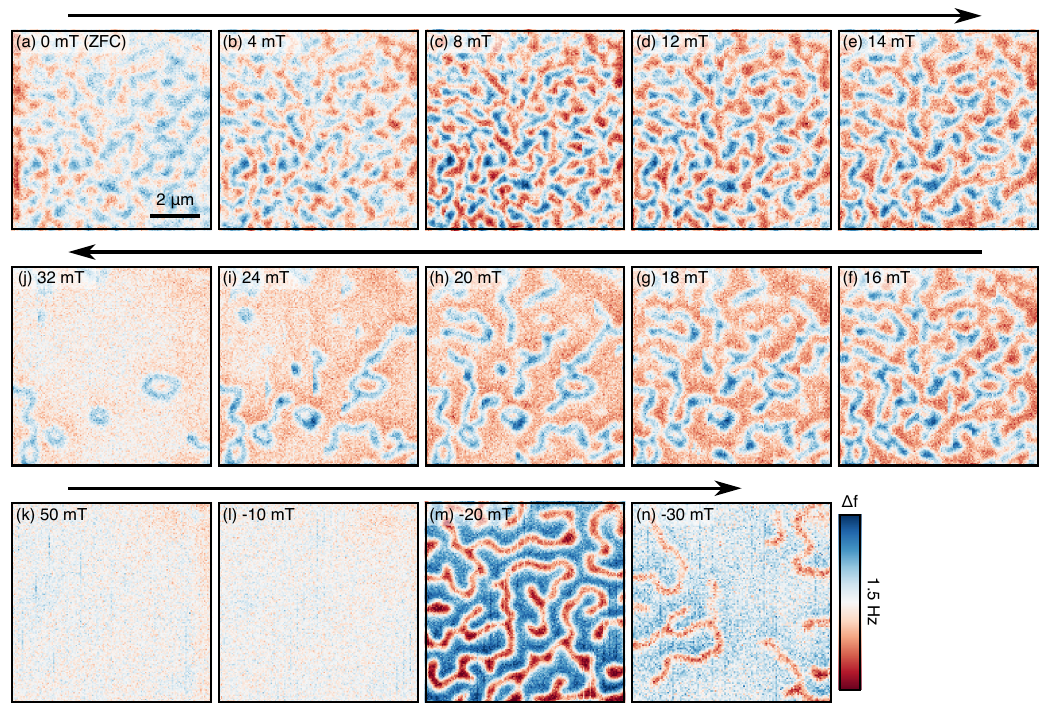}
    \caption{\label{f:full-stripes} 
        {\bf (a-n) } Constant height MFM images of the magnetic domains at the center of the MST device recorded at 5~K after zero-field cooling. Measurements were interspersed with $\Rxy$ and $\Rxx$ measurements shown in Figure~1b and in Figure~\ref{f:transport}d. The arrows indicate the order of data acquisition. The color scale on all images is 1.5 Hz, but the zero values have been offset. The tip was lifted 300~nm above the SiO$_2$ surface.
        }
\end{figure}
\clearpage

\section{Transport measurements}

\begin{figure}[h]
    \centering
    \includegraphics[width=17.78cm]{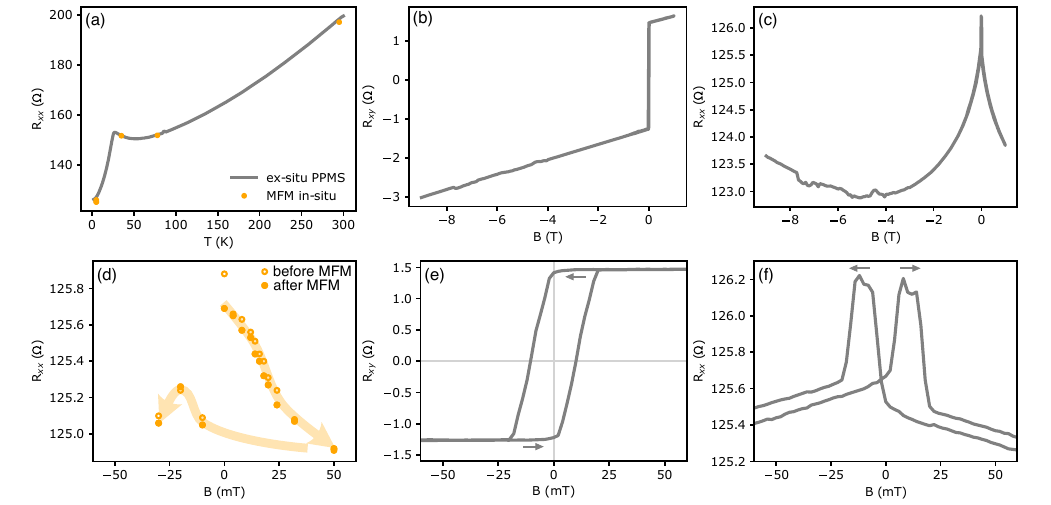}
    \caption{\label{f:transport}
        {\bf (a) } Temperature dependence of $\Rxx$, measured {\textit{ex situ}} just after sample fabrication (gray line). The orange symbols are {\textit{in situ}} measurements in the the MFM using AC source current with amplitude between 100~nA and 500~nA.
        {\bf (b-f)} Magnetic field dependence of $\Rxx$ and $\Rxy$ at low temperature measured {\textit{ex situ}} just after sample fabrication at 2~K (gray line), and {\textit{in situ}} in the MFM at 5~K (orange symbols). (e) and (f) are the same data as (b) and (c), but displayed over a smaller field range in order to make the hysteresis visible.
        {\bf (d)} AC source current amplitude: 500~nA.
        }
\end{figure}
\clearpage

\section{Modeling bubble domains}

\begin{figure}[h]
    \centering
    \includegraphics[width=17.78cm]{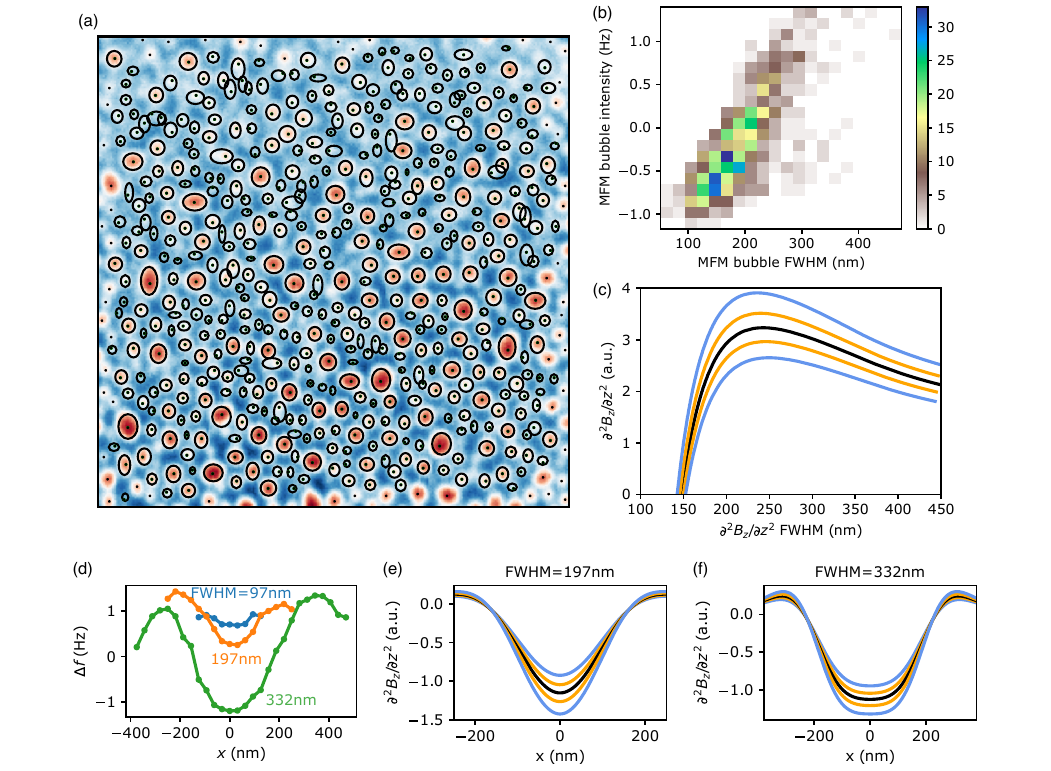}
\end{figure}
\begin{figure}[h]
    \centering
    \caption{\label{f:bubbles}
        {\bf (a) } Constant height MFM image repeated from Figure~2e, with the bubble locations and perimeters used in Figure~2k overlaid. The points show the bubble locations determined as the local minima of the image. The ellipses have major and minor axes terminated by the 4 half-minima positions in the x- and y-directions. The size of the major and minor axes are the FWHM in the x- and y-directions. Points without ellipses denote bubbles at the edge of the image for which the FWHM could not be reliably determined, and these bubbles were therefore ignored in the data analysis. While most bubbles are well-described by the ellipses, there are several where the representation is inaccurate due to the influence of neighboring bubbles on the x- and y-direction linecuts.
        {\bf (b)} Histogram of the intensities and FWHM of the bubbles in (a). The vertical and horizontal FWHM were included independently, such that each bubble is counted twice.
        {\bf (c)} Calculated $\ddBz$ peak intensity as a function of FWHM for a ferromagnetic cylindrical bubble domain. Note that the FWHM in $\ddBz$ on the x axis is related to the domain diameter by Figure~2l. The height $z$ above the ferromagnet, and the thickness $h$ of the ferromagnet are related by $z + h = 300 \textrm{nm}$ for comparison to the MFM data. The thicknesses are 
        $h= 81.2 \textrm{nm}$ (black),  
        $h = 76.7 \textrm{nm}, 85.6\textrm{nm}$ (orange), 
        $h = 71.2 \textrm{nm}, 91.4\textrm{nm}$ (blue), which correspond to the mean, 15.9, 84.1, 0.5, and 99.5 percentiles of the MST's height distribution in the approximate location of (a).
        {\bf (d)} Linecuts through 3 bubbles in the MFM image repeated from Figure~2j.
        {\bf (e, f)} Linecuts through the peak in $\ddBz$ for the same $h$ and $z$ values as in (c), and for two of the FWHM values shown in (d). The model is a good description of the mid-size bubbles (e), but the model shows a distinct flat top for large bubbles (d), which differs from the MFM data.
        Discrepancies between the model and data could arise from: (1) The size and shape of the tip's magnetic coating, (2) The presence of neighboring bubbles, (3) The finite amplitude of the tip oscillation, (4) The finite size of domain walls.
        }
\end{figure}
\clearpage

\section{Domain length scales}

\begin{figure}[h]
    \centering
    \includegraphics[width=17.78cm]{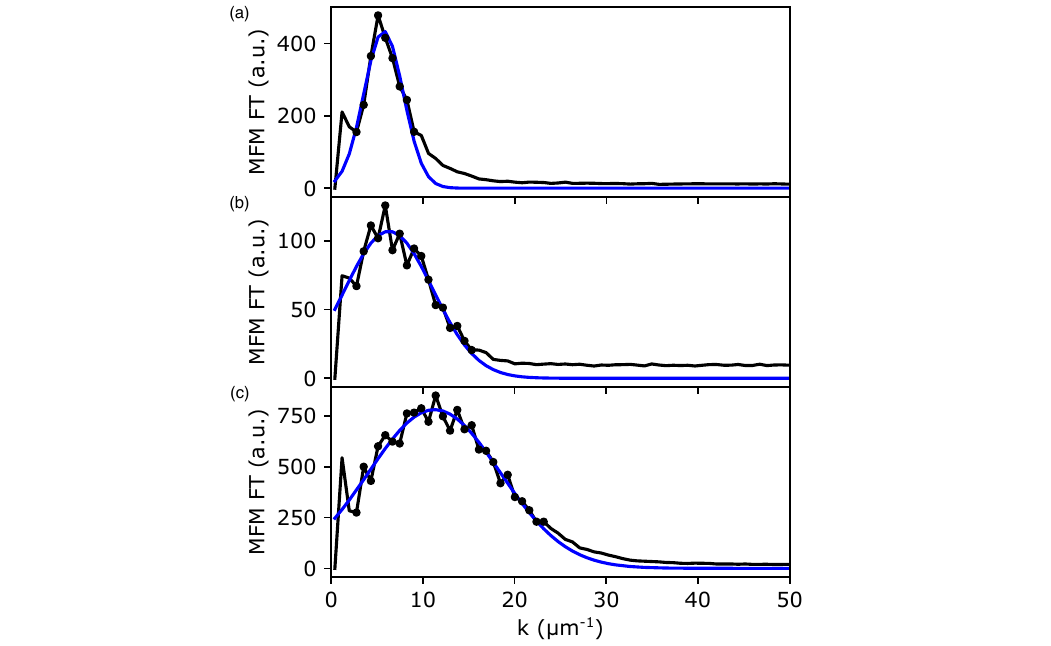}
\end{figure}
\begin{figure}[h]
    \centering
    \caption{\label{f:domain-fts}
        Angular averaged amplitude of the Fourier transforms of MFM-imaged stripe domains during magnetization reversal at $\Bext=-20\textrm{mT}$ (a) and after zero-field cooling (b), as well as MFM-imaged bubble domains after field-cooling (c). The original MFM images are Figures~1c-v, 1c-i, and 2e of the main text. The blue lines are fits to a Gaussian peak with linear background giving the peak locations and standard deviations listed in the main text.
        }
\end{figure}
\clearpage

\section{Area of stripe domains}

\begin{figure}[h]
    \centering
    \includegraphics[width=17.78cm]{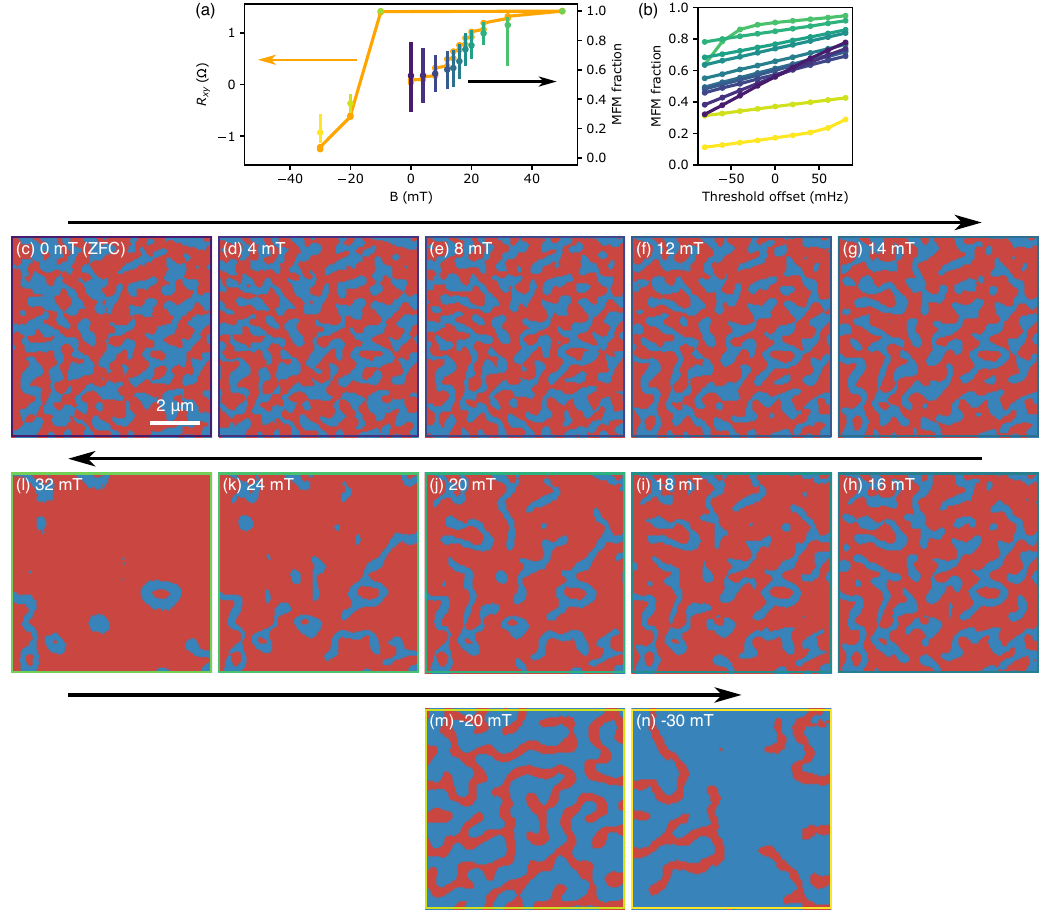}
\end{figure}
\begin{figure}[h]
    \centering
    \caption{\label{f:stripe-areas}
        {\bf(a)} Comparison of $\Rxy$ (orange) to the fraction of the MFM images covered by red domains (colored points) for the data shown in Figure~1 and Figure~\ref{f:full-stripes}. The area fractions were estimated by applying a threshold to the MFM data after subtracting a 2nd degree polynomial background and applying Gaussian smoothing ($\sigma=1$~pixel). The threshold is the midpoint between the 0.1 and 99.9 percentiles of the image. The error bars show the range of values obtained by offsetting the threshold from -80~mHz to 80~mHz. Other sources of error are not represented. For the 50~mT and -10~mT images where no domain contrast is visible, the area fractions were manually chosen to be 1.0.
        {\bf (b)} MFM area fractions calculated for a range of threshold values. Each color represents a single image, matching the points in (a).
        {\bf (c-n)} Binarized MFM images obtained by applying the threshold to calculate the points in a. The 50~mT and -10~mT images are not included because the images show no domain contrast.
        }
\end{figure}
\clearpage

\section{$\V13$ as a proxy for $\Rxy$}

\begin{figure}[h]
    \centering
    \includegraphics[width=17.78cm]{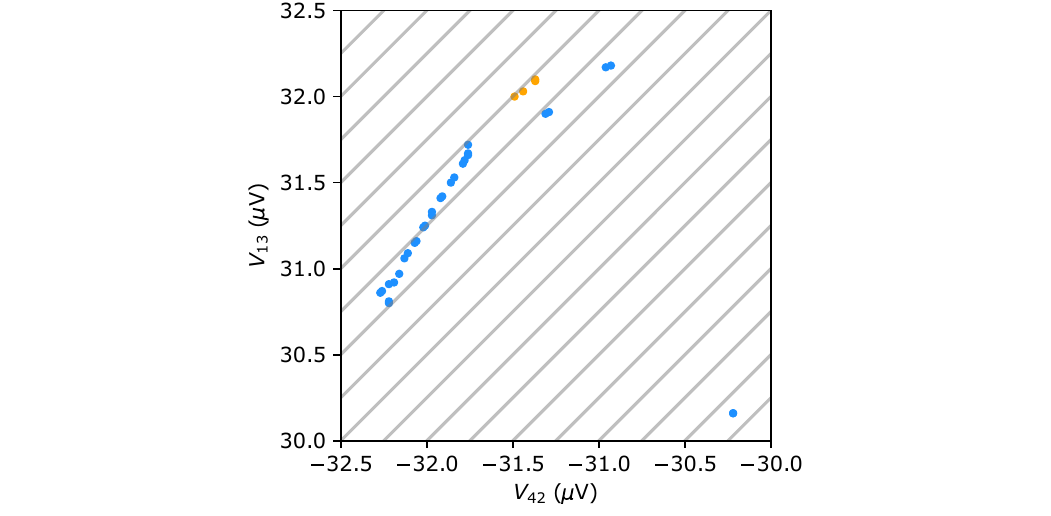}
\end{figure}
\begin{figure}[h]
    \centering
    \caption{\label{f:v13-v-rxy}
        $\V13$ and $\V42$ measurements for two 5~K data sets (blue: stripe domains under changing $\Bext$; orange: bubble domains under nominal $\Bext=0$ before and after domain manipulation with the MFM tip on 2 cools). $\Rxy$ is calculated as $\Rxy = ( \V13 + \V42 ) / 2I $, where $I$ is the source current amplitude, so changes that appear symmetrically in $\V13$ and $\V42$ are attributable to changes in $\Rxy$. The approximately linear with slope near 1 relationship between $\V13$ and $\V42$ confirms that $\V13$ is a reasonable proxy for monitoring changes in $\Rxy$ during domain manipulation. AC source current with 500~nA amplitude.
        }
\end{figure}
\clearpage

\section{Repeatability of bubble locations}

\begin{figure}[h]
    \centering
    \includegraphics[width=17.78cm]{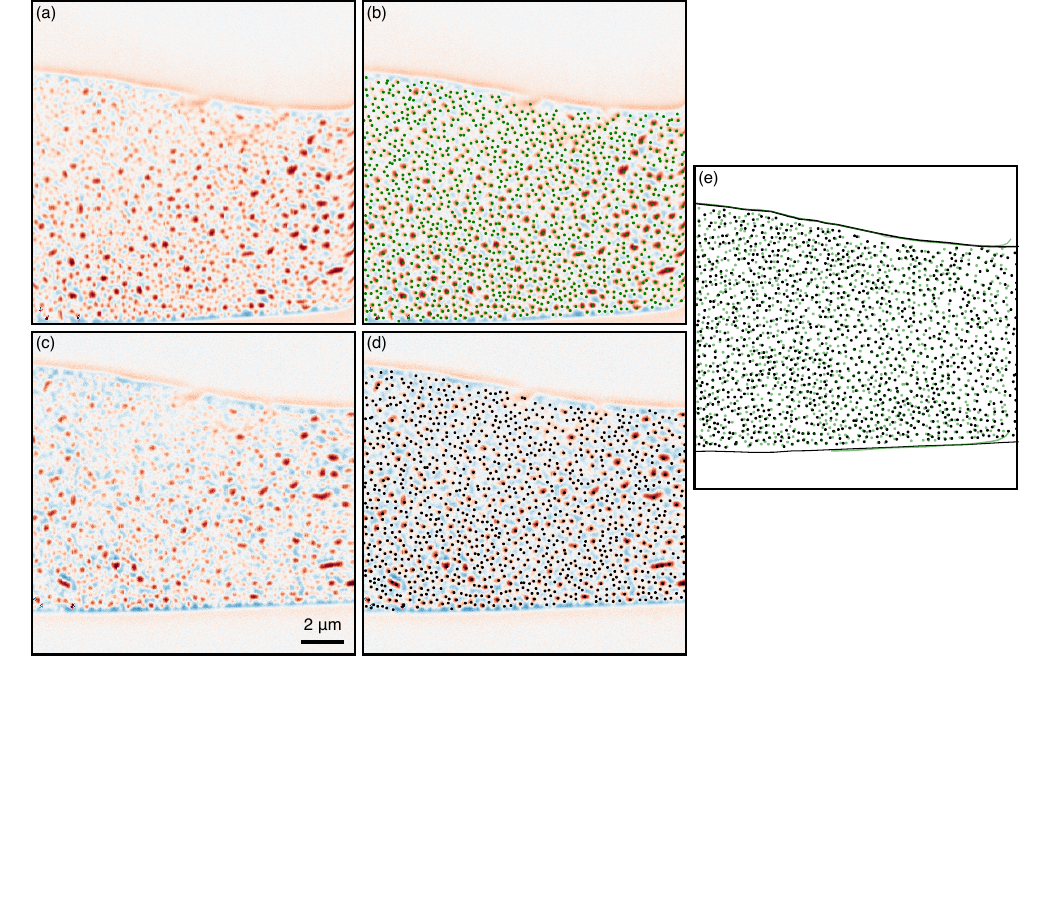}
\end{figure}
\begin{figure}[h]
    \centering
    \caption{\label{f:bubble-locations}
        To check the repeatability of the bubble domain locations, we cooled twice from 35~K to 5~K under nominal $\Bext=0$, without changing $\Bext$ in between cools, such that the residual field from the superconducting magnet should have been identical for both cools. 
        {\bf (a)} Large scale constant height MFM image of the resulting bubble domains (cool \#1). The smearing  on the right side is from piezo drift at the beginning of the image. Color scale range: 4.4~Hz.
        {\bf (b)} The bubble locations determined as the locations of all local minima of the MFM image within the area of the flake, overlaid on the MFM image.
        {(c-d)} Same as a and b, but for cool \#2. The MFM image is repeated from Figure~2a of the main text. Color scale range: 3.7~Hz.
        {\bf (e)} Bubble locations from the two cools overlaid on one another. The images have been shifted and scaled slightly in order to line up the edges of the flake from the two images (lines). Some bubble locations are identical between the cools, others do not match.
        For MFM imaging, the tip was lifted 300~nm above the SiO$_2$ surface. Temperature: 5~K.
        }
\end{figure}
\clearpage

\section{Bubble and stripe domains in bulk MST}

\begin{figure}[h]
    \centering
    \includegraphics[width=17.78cm]{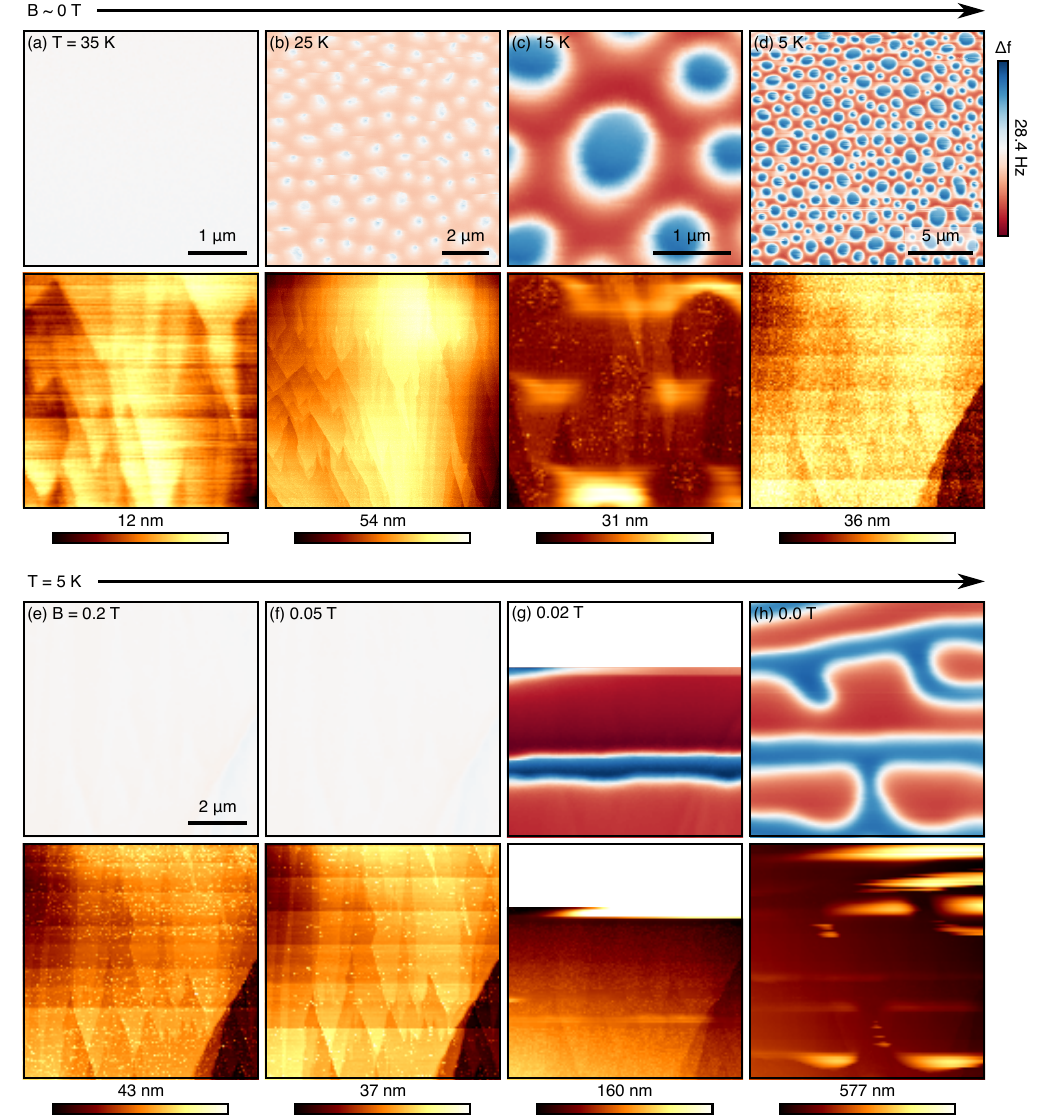}
\end{figure}
\begin{figure}[h]
    \centering
    \caption{\label{f:bulk-mst}
        {\bf (a-d)} Constant lift MFM images and their simultaneously recorded topographic images at temperatures cooling from 35~K to 5~K. No magnetic contrast is visible at 35~K, above $\Tc$. Below $\Tc$, bubble domains are observed. The topographic images show a combination of topographic features and magnetic features due to the strong tip-sample interaction.
        {\bf (e-h)} Constant lift MFM images and their simultaneously recorded topographic images at 5~K, starting at 0.2~T, where no domains are seen, and lowering to 0~T. At 0.02~T and 0~T, stripe domains are observed. Again, when domains are present, magnetic features appear in the nominally topographic images due to the strong tip-sample interaction.
        All MFM images are shown with the same color scale range 28.4~Hz, but the zero value of the images have been offset. 
        Topographic images were levelled by subtracting a planar background.
        }
\end{figure}

In this section, we show temperature-dependent and $\Bext$-dependent MFM imaging of a bulk MST crystal. We cleaved the MST crystal in air before loading into the MFM. We note several limitations of the MFM images shown here: (1) The constant lift MFM imaging technique means that MFM imaging is done after topographic scanning, so it is possible that the magnetic domains have been influenced by the tip; (2) The appearance of magnetic features in the topographic scans means that the lift height during the MFM imaging is not accurate with respect to the sample surface. Nonetheless the data shown confirms the presence of bubble domains (under small $\Bext$ cooling), and stripe domains (under $\Bext$ sweep at low temperature) in bulk MST.

\clearpage

\section{Area scaling of the Anomalous Hall Effect}

\begin{table}[h]
  \caption{\label{t:v13response} Multiple measurements of the $\V13$ response to domain writing. $s_{tip}$ is the (linear 1D) distance written by the tip. $A$ is the area of domains aligned with the tip.}
  \begin{tabular}{lclp{8cm}}
    \hline
    Experiment  & $d\V13/ds_{tip}$  & $d\V13/dA$        & Measured as\\
                & (nV/$\mu$m)       & (nV/$\mu$m$^2$) \\
    \hline
    Line domain & 8.4   &
        & Slope of $\V13$ during write\\
    Square area & 8.4   &
        & Slope of $\V13$ during the first forward scan line.\\
    Square area & 0.6   & 11 
        & $d\V13/ds_{tip}$: Peak of $\V13$ slope histogram for forward scan lines.
          $d\V13/dA = d\V13/ds_{tip} / \Delta x$, where $\Delta x$ is the pixel width, 53~nm\\
    Square area & 0.0   &
        & Peak of $\V13$ slope histogram for backward scan lines\\
    Square area &       & 12
        & Overall change in $\V13$ during write, divided by the area of red domains in Figure~4h.\\
    \hline
  \end{tabular}
\end{table}

This section provides additional details about the domain writing experiments shown in Figure~4 of the main text. We discuss: (1) additional details about the Hall response observed when writing the square area, as shown in Figure~4f-h of the main text, and (2) the consistency between multiple measurements of the Hall response to domain area.

We attempted to write a 8~$\mu$m square. The post-write imaging (Figure~4h) shows that the square was not uniformly magnetized, but instead formed a mixed domain state, with an inner blue domain inside the red domain. We examine $\V13$ recorded during the write step to understand how this pattern formed dynamically (f, g). During the write, the fast and slow scan directions are vertical and horizontal, respectively, meaning that forward (f) and backward (not shown) $\V13$ images were recorded as the tip moved up and down along each pixel before advancing one pixel at a time from left to right. The $\V13$ evolution has both smooth changes as well as abrupt jumps, which can be seen more clearly in (g) after averaging along the vertical fast scan direction. 

To quantify the write process, we extract $d\V13/ds_\mathrm{tip}$ as the slope of each vertical scan line for both forward and backward passes (Figure~4i), where $s_\mathrm{tip}$ is the linear (1D) distance written by the tip. The forward slopes are peaked around 0.6 nV/$\mu$m, while the backward slopes are peaked around 0 nV/$\mu$m, matching the expectation that the tip should typically flip the local magnetization on the forward pass and have a random influence on the backward pass. However, the forward pass of the first scan line has a much larger 
$d\V13/ds_\mathrm{tip} = 8.4 \mathrm{nV}/\mu\mathrm{m}$, which instead matches the slope seen during the line writing experiment (Figure~4d). 
The order of magnitude difference between the slope of the first and subsequent scan lines can be explained by the width being written by the tip: the first line creates a domain of a finite width on the order of hundreds of nm (Figure~4j), while subsequent scan lines advance the domain wall by one pixel width (53~nm in this case). The consistency across measurements of the $\V13$ slope when taking into account the width being written strongly supports that $\V13$ changes linearly with the area of the domains aligned with the tip. We quantify this response, $d\V13/dA$, in two ways: (1) 11 nV/$(\mu\mathrm{m})^2$, by dividing the peak slope of the forward scan lines by the pixel width, and (2) 12$\pm$1~nV/$(\mu\mathrm{m})^2$, by taking the ratio of the overall change in $\V13$ during the write to the area of red domains imaged in Figure~4h. The values of the Hall response to domain area are compiled in Table~\ref{t:v13response}.

With the relationship between $\V13$ and the sample magnetization firmly established, we return to the question of how the mixed domain state of Figure~4h formed.
If each scan line is fully polarized by the tip, then $\V13$ should change according to the linear trend shown in gray in Figure~4g.
As previously discussed, the first scan line shows a sharp drop in $\V13$ associated with writing a finite width domain. Then during the next 4 $\mu$m, $\V13$ changes largely as expected, implying that the written area is nearly fully polarized, with the exception of a few upward jumps, suggesting abrupt formation of areas anti-aligned with the tip (inner blue domains of h). During the later part of the scan frame, the $\V13$ slope shows a larger deviation from the expectation. We therefore suggest that the initial formation of the inner blue domains within the write area occurs primarily in abrupt steps. But later, the domains grow more smoothly.
There may therefore be a maximum size, on the order of a few $\mu$m$^2$ (based on the location of the first abrupt deviation) that can be uniformly polarized with our procedure at 10~K.


\bibliography{mst-domains-supp}